\documentclass[12pt]{article}
\usepackage{amssymb,amsfonts,amsmath,amscd}
\usepackage{theorem}

\newtheorem{thm}{\textbf{Theorem}}
\newtheorem{prop}{\textbf{Proposition}}

\newtheorem{lem}{\textbf{Lemma}}
\newtheorem{rema}{R e m a r k}

\begin{document}

\title{Universality at the Edge for Unitary Matrix Models}

\author{M. Poplavskyi\\ Mathematics Division, 
\\B. Verkin Institute for Low
Temperature Physics and Engineering\\
 National Academy of Sciences of Ukraine\\
47 Lenin Ave., Kharkiv 61103, Ukraine\\
\smallskip {\rm E-mail: poplavskiymihail@rambler.ru}}

\date{Received August 5, 2012}

%%%%%%%%%%%%%%%%%%%%%%%%
\protect\maketitle %\markright{}
%%%%%%%%%%%%%%%%%%%%%%%%
\def\thefootnote{\ifcase\value{footnote}\or {}\or ${\star}$\or ${\star\star}$\or ${\star\!\star\!\star\!}$ \fi}
%%%%%%%%%%%%%%%%%%%%%%%%
\newenvironment{proof}{\begin{trivlist}\item[]
{\quad\, P r o o f. }}{\hfill\rule{0.5em}{0.5em}\end{trivlist}}
%%%%%%%%%%%%%%%%%%%%%%%%%
\renewcommand{\thesubsection}{\arabic{subsection}.}
\renewcommand{\theequation}{\thesubsection\arabic{equation}}
\renewcommand{\thethm}{\thesubsection\arabic{thm}}
\renewcommand{\thelem}{\thesubsection\arabic{lem}}
\renewcommand{\theprop}{\thesubsection\arabic{prop}}
\renewcommand{\therema}{\thesubsection\arabic{rema}}
%%%%%%%%%%%%%%%%%%%%%%%%%
%\numberwithin{equation}{section}
%\numberwithin{theorem}{section}
%%%%%%%%%%%%%%%%%%%%%%%%

%dislpay style definitions
\def\dfrac{\displaystyle\frac}
\def\sumd{\displaystyle\sum}
\def\limd{\displaystyle\lim}
\def\intd{\displaystyle\int}
\def\iintd{\displaystyle\iint}
\def\ointd{\displaystyle\oint}
\def\prodd{\displaystyle\prod}

%matrices definitions
\def\G{G^{\left(n\right)}}
\def\g{g^{\left(n\right)}}
\def\M{M^{\left(n\right)}}
\def\L{L^{\left(n\right)}}
\def\C{C^{\left(n\right)}}
\def\CRm{C^{\left(n\right)}_{r_{-}}}
\def\CRp{C^{\left(n\right)}_{r_{+}}}
\def\CRpm{C^{\left(n\right)}_{r_{\pm}}}
\def\Th{\Theta^{\left(n\right)}}
\def\Res{R^{\left(n\right)}}
\def\U{U^{\left(n\right)}}
\def\V{V^{\left(n\right)}}
\def\Rstar{R^{\star}}
\def\rstar{r^{\star}}
\def\D{D^{\left(n\right)}}

%OPUC definitions
\def\wnl{w_n\left(\lambda\right)}
\def\wnm{w_n\left(\mu\right)}
\def\K{K^{\left(n\right)}}
\def\P{P^{\left(n\right)}}
\def\Q{Q^{\left(n\right)}}
\def\Ps{P^{* \left(n\right)}}
\def\Qs{Q^{* \left(n\right)}}
\def\X{\chi^{\left(n\right)}}
\def\Xh{\widehat{\chi}^{\left(n\right)}}
\def\rhon{\rho^{\left(n\right)}}
\def\alphan{\alpha^{\left(n\right)}}
\def\sn{s^{\left(n\right)}}

%remainders
\def\Obig{\underline{O}}
\def\Osmall{\overline{o}}

%brackets definitions
\def\BrLambda{\left(\lambda\right)}
\def\BrTheta{\left(\theta\right)}
\def\BrMu{\left(\mu\right)}
\def\BrLambdaMu{\left(\lambda,\mu\right)}
\def\BrLambdaZ{\left(\lambda,z\right)}
\def\BrMuW{\left(\mu,w\right)}
\def\BrLambdaMinMu{\left(\lambda - \mu\right)}
\def\BrZ{\left(z\right)}
\def\BrW{\left(w\right)}
\def\BrZW{\left(z,w\right)}
\def\BrXY{\left(x,y\right)}
\def\BrXZeta{\left(x,\zeta\right)}
\def\BrYZeta{\left(y,\zeta\right)}
\def\BrYKsi{\left(y,\xi\right)}
\def\BrZeta{\left(\zeta\right)}
\def\BrZetaKsi{\left(\zeta,\xi\right)}

%constants n,k
\def\epsnk{\varepsilon_{n,k}}
\def\Xn{X^{\left(n\right)}}
\def\Yn{Y^{\left(n\right)}}
\def\xn{x^{\left(n\right)}}
\def\yn{y^{\left(n\right)}}
\def\dn{\delta^{\left(n\right)}}
\def\ddn{\widehat{\delta}^{\left(n\right)}}
\def\Dn{\Delta^{\left(n\right)}}
\def\rn{r^{\left(n\right)}}
\def\dnk{d_{n+k}}
\def\ndnk{d_{n+k+1}}

%constants theta
\def\p{p_\theta}
\def\c{c_\theta}
\def\st{\sin\theta}
\def\ct{\cos\theta}
\def\hth{\dfrac{\theta}{2}}
\def\sht{\sin\dfrac{\theta}{2}}
\def\cht{\cos\dfrac{\theta}{2}}
\def\tht{\tan\dfrac{\theta}{2}}
\def\ctht{\cot\dfrac{\theta}{2}}
\def\ssht{\sin^2\dfrac{\theta}{2}}
\def\ccht{\cos^2\dfrac{\theta}{2}}
\def\ttht{\tan^2\dfrac{\theta}{2}}
\def\ctctht{\cot^2\dfrac{\theta}{2}}

%functions definitions
\def\e{e^}
\def\Fn{\mathcal{F}_n}
\def\Kn{\mathcal{K}_n}
\def\f{f^{\left(n\right)}}
\def\QA{Q_{Ai}}
\def\AFrac{\dfrac{Ai\left(x\right)Ai'\left(y\right) -
Ai'\left(x\right)Ai\left(y\right)}{x-y}}
\def\FA{F_{Ai}}

%Airy stuff
\def\Lf{\mathcal{L}}
\def\Rz{\mathcal{R}_\zeta}
\def\Rzc{\mathcal{R}_\overline{\zeta}}

%%%%%%%%%%%%%%%%%%%%%%%%
\begin{abstract}

Using the results on the $1/n$-expansion of the Verblunsky coefficients for
a class of polynomials orthogonal on the unit circle with $n$ varying
weight, we prove that the local eigenvalue statistic for unitary matrix
models is independent of the form of the potential, determining the matrix
model. Our proof is applicable to the case of four times differentiable
potentials and of supports, consisting of one interval. \vskip2mm

\smallskip


\end{abstract}\smallskip%\medskip

%%%%%%%%%%%%%%%%%%%%%%%%%%%%%%%%%%%\begin{document}

\setcounter{subsection}{0}\setcounter{thm}{4}\setcounter{prop}{0}\setcounter{rema}{1}

\begin{center}
\subsection{Introduction}
\end{center}

We study a class of random matrix ensembles known as unitary matrix models.
These models are defined by the probability law

\begin{equation}\label{d:Model}
p_n\left(U\right)d\mu_n\left(U\right)=Z_{n,2}^{-1} \exp
\left\lbrace -n \hbox{Tr} V\left( \dfrac{U+U^*}{2} \right)
\right\rbrace d\mu_n\left(U\right),
\end{equation}

\noindent
where $U=\{U_{jk}\}_{j,k=1}^n$ is an $n \times n$
unitary matrix, $\mu_n\left(U\right)$ is the Haar measure on the
group $U(n)$, $Z_{n,2}$ is the normalization constant, and
$V:[-1,1] \rightarrow \mathbb{R}$ is a continuous function
called the potential of the model.
Denote $e^{i \lambda_j}$ the eigenvalues of the unitary matrix $U$.
The joint probability density of $\lambda_j$, corresponding to
\eqref{d:Model}, is given by (see \cite{Me:91})

\begin{equation}\label{d:jd}
p_n \left( \lambda_1, \ldots , \lambda_
n\right)=\dfrac{1}{Z_n}\prodd\limits_{1 \leq j < k \leq n}
\left| e^{i \lambda_j} -  e^{i \lambda_k}\right|^2
\exp
\left\lbrace
 -n \sum\limits_{j=1}^{n} V \left( \cos\lambda_j
\right)
\right\rbrace.
\end{equation}

Normalized Counting Measure of eigenvalues (NCM) is given by

\begin{equation*}
N_n\left(\Delta\right)=n^{-1}\sharp \left\{\lambda_l^{(n)} \in
\Delta, \, l=1,\ldots,n\right\}, \quad \Delta \subset [-\pi,\pi].
\end{equation*}

The random matrix theory deals with several asymptotic regimes of
the eigenvalue distribution. The global regime is centred around the
weak convergence of NCM. It is well known (see e.g. \cite{Kol:97})
that for some smooth conditions for the potential $V$ there exists a measure
$N \in \mathcal{M}_1
\left([-\pi,\pi]\right)$ with a compact support $\sigma$ such that
$N_n$ converges to $N$ in probability .

Let
\begin{equation*}
p_l^{\left( n \right)} \left( \lambda_1, \ldots, \lambda_l \right)
= \intd p_n \left( \lambda_1, \ldots, \lambda_l , \lambda_{l+1}
,\ldots ,\lambda_n\right) \, d\lambda_{l+1} \ldots d\lambda_n
\end{equation*}
be the $l$~-th marginal density of $p_n$. The local
regime of eigenvalue distribution describes the asymptotic
behaviour of marginal densities when their arguments are on the
distances of order of the typical distance between eigenvalues.
The universality conjecture of marginal densities was suggested by
Dyson (see \cite{Dy:62}) in the early 60s. He supposed that their
asymptotic behaviour depends only on the ensemble symmetry group
and does not depend on other ensemble parameters. First rigorous
proofs for the hermitian matrix models with non-quadratic $V$
appeared only in the 90s. The case of general $V$ which is locally
$C^3$ function was studied in \cite{Pa-Sh:97}. The case of real
analytic $V$ was studied in \cite{De:99}, where the asymptotic
behaviour of orthogonal polynomials was obtained. For the unitary
matrix models the bulk universality was proved for $V=0$ (see
\cite{Dy:62}), and for the locally $C^3$ functions (see
\cite{Po:08}). The edge universality was proved only in the case
of the linear $V$ (see \cite{Jo-D:98}). In the present paper we
prove the universality conjecture for UMM with a smooth potential
$V$ in the case of one-interval support $\sigma$ of the limiting
NCM.

It was proved in \cite{Kol:97} that the limiting measure can be obtained as a
unique minimizer of the functional
\begin{equation*}
\mathcal{E}[m]=\intd\limits_{-\pi}^{\pi} V(\cos \lambda
)m(d\lambda )- \intd\limits_{-\pi}^{\pi} \log \left|e^{i \lambda}
- e^{i\mu} \right|m(d\lambda )m(d\mu )
\end{equation*}
in the class of unit measures on the interval $\left[-\pi,\pi\right]$ (see
\cite{ST:97} for the existence and properties of the solution). It is well
known, in particular, that for smooth $V'$ the equilibrium measure has a
density $\rho$ which is uniquely defined by the condition that the function
\begin{equation}\label{d:u}
u\left(\lambda\right)=
V\left(\cos\lambda\right)-2\intd\limits_{\sigma} \log
\left|e^{i\lambda}-e^{i\mu}\right| \rho\left(\mu\right) d\mu
\end{equation}
takes its minimum value if $\lambda \in \sigma = \mathrm{supp} \,  \rho$.
From this condition in the case of differentiable $V$ one can obtain the
following integral equation for the equilibrium density $\rho$:
\begin{equation}\label{e:V_Rho}
\left( V \left( \cos \lambda \right) \right)'=
v.p.\intd\limits_{\sigma} \cot
\dfrac{\lambda-\mu}{2} \rho \left(\mu\right) d\mu, \quad
\mbox{for}\, \lambda \, \in \, \sigma.
\end{equation}

We also use the weak convergence of the first marginal density
$\rho_n\BrLambda = p_1^{(n)}$ proved in \cite{Kol:97}.\medskip

\begin{prop}\label{p:wconv}
For any $\phi \in H^1 \left(-\pi,\pi\right),$

\begin{equation}\label{e:wconv}
\left|%
\intd\phi\left(\lambda\right)\rho_n\left(\lambda\right)%
\,d \lambda%
-%
\intd\phi\left(\lambda\right)\rho
\left(\lambda\right)%
\,d \lambda%
\right|%
\leq%
C \left\| \phi \right\|^{1/2}_{2} \left\| \phi' \right\|^{1/2}_{2}
n^{-1/2} \ln^{1/2} n,
\end{equation}
where $\left\|\cdot\right\|_2$ denotes $L_2$ norm on $[-\pi,\pi]$.
\end{prop}
We consider here the case of one interval $\sigma$. Our main conditions on
the potential $V$ are\medskip

\textbf{Condition C1.} \textit{ The support $\sigma$ of the equilibrium
measure is a single symmetric subinterval of the interval
$\left[-\pi,\pi\right]$, i.e.,
\begin{equation*}
\sigma = \left[-\theta,\theta\right], \, \text{with} \quad \theta < \pi.
\end{equation*}}
\begin{rema}
{\rm In fact, there is one more possibility to have one-interval $\sigma$.
 Another case is some left symmetric arc of the circle, i.e.,
 $\left[\pi-\theta,\pi+\theta\right]$.
In this case we replace $V\left(\cos x\right)$ in \eqref{d:jd} by
$V\left(\cos\left(\pi-x\right)\right)$. This replacement will rotate all
eigenvalues on the angle $\pi$ and we will have the support from condition
C1.}
\end{rema}

\textbf{Condition C2.} \textit{ The equilibrium density $\rho$ has no zeros
in $(-\theta,\theta)$ and
\begin{equation*}
\rho\left(\lambda\right) \sim C  \left|\lambda \mp \theta\right|^{1/2},
\, \text{for}\; \lambda \rightarrow \pm
\theta,
\end{equation*}
\noindent and the function $u\left(\lambda\right)$ of \eqref{d:u} attains
its minimum if and only if $\lambda$ belongs to~$\sigma$. }

\begin{rema}
{\rm From this condition we obtain the necessary scaling for marginal
densities at the edge of $\sigma$
\begin{equation}
\intd\limits_{\Delta} \rho \left( \lambda \right) d\lambda \sim n^{-1} \Rightarrow
\mid
\Delta
\mid \sim n^{-2/3},
\end{equation}}
\end{rema}
\noindent hence the typical distance between eigenvalues is of order
$n^{-2/3}$.\medskip

\textbf{Condition C3.} \textit{
 $V\left(\cos \lambda\right)$ possesses four bounded derivatives
 on $\sigma_{\varepsilon}=\left[-\theta-\varepsilon,
\theta+\varepsilon\right]$.
}

The following simple representation of $\rho$ plays an important role in
our asymptotic analysis (see \cite{Po:12})\medskip\setcounter{prop}{3}

\begin{prop}\label{p:Rho}
Under conditions C1-C3 the density $\rho$ has the form
\begin{equation*}
\rho \left(\lambda\right) = \dfrac{1}{4\pi^2}\chi
\left(\lambda\right)P\left(\lambda\right)\mathbf{1_{\sigma}},
\end{equation*}
where
\begin{equation}\label{d:P}
\chi \left(\lambda\right) =
\sqrt{\left|\cos\lambda-\cos\theta\right|}, \quad
P\left(\lambda\right) = \intd\limits_{-\theta}^{\theta} \dfrac{\left(V
\left(\cos \mu\right)\right)' - \left(V \left(\cos
\lambda\right)\right)'}{\sin\left(\mu-\lambda\right)/2}
\dfrac{d\mu}{\chi \left(\mu\right)}.
\end{equation}
\end{prop}

The main result of the paper is the following theorem\medskip

\begin{thm}\label{t:main}
Consider the unitary matrix ensemble of the form \eqref{d:Model},
sa\-tis\-fying conditions {\rm C1--C3} above. Then
\begin{itemize}
\item
for the endpoints $\theta_{\pm} = \pm \theta$
and any positive integer $l$ the rescaled marginal density
\begin{equation}\label{e:md_r}
\left( \gamma n^{2/3} \right)^{-l}
\dfrac{n!}{\left( n - l\right)! }
p_l^{\left( n\right) }
\left(
\theta_{\pm} \pm t_1/\gamma n^{2/3},
\ldots,
\theta_{\pm} \pm t_l/\gamma n^{2/3}
\right)
\end{equation}
with the sign $\pm$ corresponding to $\theta_{\pm}$ and
\begin{equation*}
\gamma = \tan^{1/3} \theta/2 \left(\dfrac{P \BrTheta}{4\pi}\right)^{2/3}
\end{equation*}
converges weakly, as $n \to \infty$, to $ \det \left\lbrace \QA \left(
t_j,t_k\right) \right\rbrace_{j,k = 1}^{l}, $ where $\QA\BrXY $ is the
Airy kernel
\begin{equation}\label{d:AiKer}
\QA\BrXY=\dfrac{
Ai\left( x\right)Ai'\left( y\right)-
Ai'\left(x \right)Ai\left(y \right) }{x-y};
\end{equation}
\item if $\Delta \subset \mathbb{R}$ is a finite union of disjoint
    bounded intervals and
\begin{equation*}
E_n\left( \Delta_n\right) = \mathbb{P} \left( \Delta_n \text {does
not contain eigenvalues of } U\right)
\end{equation*}
is the hole probability for $\Delta_n = \theta_{\pm} \pm \Delta/\gamma
n^{2/3}$, then
\begin{equation}\label{e:Fred}
\lim_{n \to \infty} E_n\left( \Delta_n \right)
= 1 + \sumd_{l=1}^{\infty}
\dfrac{\left(-1 \right)^l}{l!}
\intd\limits_{\Delta} dt_1 \ldots dt_l
\det \left\lbrace
\mathcal{K}
\left( t_j,t_k\right)
\right\rbrace_{j,k = 1}^{l},
\end{equation}
i.e., the limit is the Fredholm determinant of the integral operator
$\mathcal{K}_{\Delta}$ defined by the kernel $\mathcal{K}$ on the set
$\Delta$.
\end{itemize}

\end{thm}

The paper is organized as follows. In Section~2 %\ref{s:opuc}
we give a brief outline of the orthogonal polynomials  method. In Section~3
%\ref{s:main}
we prove the main Theorem~\ref{t:main} using some technical
results. These results are proved in
Section~\ref{s:aux}\medskip\setcounter{equation}{0}
\setcounter{equation}{0}

{\centering\subsection{Orthogonal Polynomials}\label{s:opuc}}

We prove Theorem~\ref{t:main}, using the orthogonal polynomials technique.
This method is based on a simple observation. Joint eigenvalue distribution
\eqref{d:jd} is expressed in terms of the Vandermonde determinant of powers
of $\e{i\lambda_k}$, and therefore by the properties of determinants, can
be written in terms of the determinant of any system of linearly
independent trigonometric polynomials. We consider a system of polynomials
orthogonal on the unit circle(OPUC) with a varying weight. Let

\begin{equation*}
w_n \left(\lambda\right) = \e{-nV\left(\cos\lambda\right)}
\end{equation*}
be the weight function for the system of polynomials. Then the system can
be obtained from $\left\{e^{i k \lambda}\right\}_{k=0}^{\infty}$ if we use
the Gram-Schmidt procedure in $ L^{\left(n\right)}:=L_2\left(
\left[-\pi,\pi\right], w_n\left(\lambda\right) \right)$ with the inner
product
\begin{equation*}
\left\langle f,g\right\rangle_n=\displaystyle\intd\limits_{-\pi}^{\pi}
f\left(x\right) \overline{g\left(x\right)} w_n\left(x\right)dx.
\end{equation*}

Hence, for any $n$ we get the system of trigonometric polynomials
 $\left\lbrace\P_k\BrLambda\right\rbrace_{k=0}^{\infty}$
which are orthonormal in $L^{\left(n\right)}$. One can see from
the Szeg$\ddot{o}$'s condition that the system
$\left\{\P_k\left(\lambda\right)\right\}_{k=0}^{\infty}$ is not
complete in $L^{\left(n\right)}$. To construct the complete system
one should also include  polynomials with respect to
$e^{-i\lambda}$. Thus, following \cite{CMV:03}, we introduce the
Laurent polynomials
\begin{equation}\label{d:X}
\begin{array}{lcl}
\chi_{2k}^{\left(n\right)}\left(\lambda\right) &=&
e^{ik\lambda}\P_{2k}\left(-\lambda\right),
\\
\chi_{2k+1}^{\left(n\right)}\left(\lambda\right) &=&
e^{-ik\lambda}\P_{2k+1}\BrLambda.
\end{array}
\end{equation}

 It is easy to check (see, e.g., \cite{CMV:03, Si:05}) that the
system $\left\{\chi_k^{\left(n\right)}\BrLambda\right\}_{k=0}^{\infty}$ is
an orthonormal basis in $L^{\left(n\right)}$.  Moreover, it was proved in
\cite{CMV:03} that the functions $\chi_k^{\left(n\right)}$ satisfy some
five term recurrent relations. Let $\alpha_k^{\left(n\right)}$ and
$\rho_k^{\left(n\right)}$ be the Verblunsky coefficients of the system
$\left\{\X_k\left(\lambda\right)\right\}_{k=0}^{\infty}$ (for the
definition and properties see \cite{Po:12}). Denote by
\begin{equation*}
\Theta^{\left(n\right)}_j = \left(
\begin{array}{cc}
-\alpha^{\left(n\right)}_j & \rho^{\left(n\right)}_j \\
\rho^{\left(n\right)}_j & \alpha^{\left(n\right)}_j%
\end{array}
\right),
\end{equation*}
\begin{equation*}
M^{\left(n\right)} = E_1 \oplus \Theta^{\left(n\right)}_2 \oplus
\Theta^{\left(n\right)}_4 \oplus ..., \quad L^{\left(n\right)} =
\Theta^{\left(n\right)}_1 \oplus \Theta^{\left(n\right)}_3 \oplus
\Theta^{\left(n\right)}_5 \oplus ...,
\end{equation*}
\begin{equation}\label{e:C_ML}
C^{\left(n\right)} = M^{\left(n\right)}L^{\left(n\right)} .
\end{equation}

From the properties of the Verblunsky coefficients one can see that the
semi-infinite matrices $\M$ and $\L$ are symmetric, three diagonal and
unitary. $C^{\left(n\right)}$ is also a unitary five diagonal matrix.
Finally, using the above notations, we can write the recurrence relations
as
\begin{equation*}
\quad e^{i\lambda}\overrightarrow{%
\X} = C^{\left(n\right)} \overrightarrow{%
\X}.
\end{equation*}
Hence, $\C$ is a matrix presentation of the multiplication operator
by $\e{i\lambda}$ in the basis
$\left\{\X_k\left(\lambda\right)\right\}_{k=0}^{\infty}$.

The main advantage of the orthogonal polynomials technique is the
determinant formulas which can be obtained in the same way as in
\cite{Me:91},
\begin{equation}\label{e:Det}
\dfrac{n!}{\left( n -l\right)! }
p_l^{\left( n\right) }
\left(
\lambda_1,
\ldots,
\lambda_l
\right)
=
\det \left\{\K_{n}\left(\lambda_j,\lambda_k\right)\right\}_{j,k=1}^{l},
\end{equation}
where
\begin{equation}\label{d:K}
\K_m\BrLambdaMu = \sumd\limits_{k=0}^{m-1} \X_k\BrLambda \overline{\X_k\BrMu}
w_n^{1/2}\BrLambda w_n^{1/2}\BrMu
\end{equation}
is the reproducing kernel of the system
$\left\{\X_k\left(\lambda\right)\right\}_{k=0}^{\infty}$. Similarly to
\cite{Pa-Sh:03}, the weak convergence of the kernel $\K_n$ to $\mathcal{K}$
as $n \to \infty$ will prove
Theorem~\ref{t:main}.

\bigskip

\setcounter{equation}{0}\setcounter{lem}{0}\setcounter{prop}{1}

{\centering\subsection{Proof of Theorem~\ref{t:main}}\label{s:main}}

To prove the weak convergence of the reproducing kernel \eqref{d:K}, we use
the lemma (see \cite{Pa-Sh:03})\medskip

\begin{lem}\label{l:ker}
Consider the sequence of functions $\mathcal{K}_n : \mathbb{R}
\times \mathbb{R} \to \mathbb{C}$ and define for $\Im \zeta,\xi
\neq 0,$
\begin{equation}\label{d:Fn}
\Fn \BrZetaKsi = \iintd
\Im \dfrac{1}{x-\zeta}
\Im \dfrac{1}{y-\xi}
\left|\mathcal{K}_{n}\BrXY\right|^2
d x
d y.
\end{equation}
Assume that there exists $\mathcal{F} \BrZetaKsi$ of the form
\begin{equation}\label{d:KA}
\mathcal{F} \BrZetaKsi = \iintd
\Im \dfrac{1}{x-\zeta}
\Im \dfrac{1}{y-\xi}
\left|\mathcal{K}\BrXY\right|^2
d x
d y,
\end{equation}
with $\mathcal{K}$ bounded uniformly in each compact in $\mathbb{R}^2$ and
such that for any fixed $A>0$ uniformly on the set
\begin{equation}\label{d:On}
\Omega_A = \left\lbrace\zeta,\xi :
1 \leq \Im \zeta,\Im \xi \leq A,
\left|\Re \zeta,\Re \xi\right|\leq A \right\rbrace
\end{equation}
we have
\begin{equation}\label{e:Fn_conj}
\left| \Fn\BrZetaKsi - \mathcal{F}\BrZetaKsi\right|
\leq \varepsilon_n, \quad \varepsilon_n \to 0,
\text{as } n \to \infty.
\end{equation}
Then for any intervals $I_1, I_2 \subset \mathbb{R}$
\begin{equation*}
\lim_{n \to \infty}
\intd\limits_{I_1} dx \intd\limits_{I_2}  dy
\left|
\mathcal{K}_{n}\BrXY\right|^2
=
\intd\limits_{I_1} d x \intd\limits_{I_2}  d y
\left|
\mathcal{K}\BrXY\right|^2.
\end{equation*}
\end{lem}

The lemma helps to prove the convergence of $\left|\mathcal{K}_n\right|^2$
to $\left|\mathcal{K}\right|^2$. Similarly, we can check the convergence of
$\mathcal{K}_n\left(t_1,t_2\right)\mathcal{K}_n\left(t_2,t_3\right)
\ldots\mathcal{K}_n\left(t_l,t_1\right)$ for any $l \in \mathbb{N}$. To
prove the second part of Theorem~\ref{t:main}, we use another proposition
from \cite{Pa-Sh:03}.\medskip

\begin{prop}\label{p:Fred}
Let $\Delta \subset \mathbb{R}$ be a system of disjoint intervals as
in Theorem~\ref{t:main} and let $\mathcal{K}_n : L_2 \left(\Delta\right) \to
L_2 \left(\Delta\right)$ be a sequence of positive definite integral
operators with kernels $\mathcal{K}_n\BrXY$ and
$\mathcal{K} : L_2 \left(\Delta\right) \to
L_2 \left(\Delta\right)$ a positive definite integral
operator with kernel $\mathcal{K}\BrXY$, such that for any $l \in \mathbb{N}$,
$\det \left\lbrace \mathcal{K}_n \left(t_j,t_k\right) \right\rbrace_{j,k=1}^l
\to \det \left\lbrace \mathcal{K} \left(t_j,t_k\right) \right\rbrace_{j,k=1}^l$
weakly as $n\to \infty$. Assume also that for any $\Delta$ there exists
$C_{\Delta}$ such that
\begin{equation}\label{e:Fr_conj}
\intd\limits_{\Delta} \mathcal{K}_n \left(s,s\right) ds
\leq C_{\Delta}.
\end{equation}
Then,  for the Fredholm determinants of $\mathcal{K}_n$ and $\mathcal{K}$
we have
\begin{equation*}
\lim\limits_{n \to \infty} \det \left(1 - \mathcal{K}_n\right)
= \det \left(1 - \mathcal{K}\right).
\end{equation*}
\end{prop}

We are going to use Lemma~\ref{l:ker} for the scaled reproducing kernel of
the system of OPUC. Let
\begin{equation}\label{d:Kn}
\Kn\BrXY = n^{-2/3} \K_n
\left(\theta+xn^{-2/3},\theta+yn^{-2/3}\right)
\mathrm{1}_{\left|x,y\right|\leq \c n^{2/3}}
\end{equation}
for some small enough $\theta$-dependent constant $\c$. This will be
sufficient in view of the following lemma (the analogue of Theorem 11.1.4,
\cite{Pa-Sh:b})\setcounter{lem}{2}\setcounter{prop}{3}\smallskip

\begin{lem}\label{l:Kn_int}
Let the model \eqref{d:Model} satisfy conditions C1-C3. Then, for
any $n$-independent $\varepsilon > 0,$ there exists a constant
$d_{\varepsilon}>0$ such that
\begin{equation*}
\intd\limits_{\sigma_{\varepsilon}^c}
\K_n\left(\lambda,\lambda\right)d\lambda \leq C \e{-nd_{\varepsilon}}.
\end{equation*}
\end{lem}

Since the polynomials $\X_k$ are functions of $\e{i\lambda}$, it is more
convenient to define a little bit different from  \eqref{d:Fn}
transformation and estimate the difference between it and \eqref{d:Fn}.
Hence, we consider the following transformation:
\begin{equation}\label{d:Fn2}
F_n \BrZW = n^{-4/3}
\iintd\limits_{\left[-\pi,\pi\right]}
G\left(z-\lambda\right)
G\left(w-\mu\right)
\left|\K_n \BrLambdaMu\right|^2 d\lambda d\mu,
\end{equation}
with
\begin{equation}\label{d:G}
G\BrZ = \Re g\BrZ, \mbox{and} \, g\BrZ = \dfrac{1+\e{iz}}{1-\e{iz}}
\end{equation}
being the analogues of the Poisson and the Herglotz
transformations.\smallskip%\medskip

\begin{prop}\label{p:G_prop}
It follows from the definition of $g\BrZ$ that
\begin{equation*}
g\BrZ = i \cot \dfrac{z}{2},
\quad
g\left(z-\lambda\right) =
\dfrac{\e{i\lambda} + \e{iz}}
{\e{i\lambda} - \e{iz}}.
\end{equation*}
For $z=x+iy$ we have $ g\left(x+iy\right) = \dfrac{i\sin x + \sinh y}{\cosh
y - \cos x}$, hence $ \overline{g\BrZ} = -g\left(\overline{z}\right) $. And
for $G\BrZ$ we get
\begin{equation*}
G\left(x+iy\right)=
\dfrac{\sinh y}{\cosh y - \cos x},
\quad
G\left(z-\lambda\right) = \Im \cot \dfrac{\lambda-z}{2}.
\end{equation*}
Moreover, $G\BrZ$ is a Nevanlinna function and
\begin{equation}\label{e:gsq}
\left|
g\BrZ
\right|^2 = -1 + 2\coth \Im z \cdot G\BrZ.
\end{equation}
\end{prop}

The difference between the new transformation and the old one can be
estimated in the following way:\smallskip%\medskip

\begin{prop}\label{p:G_est}
Let $z=\theta+\zeta n^{-2/3}$ and $w=\theta+\xi n^{-2/3}$ with
$\left|\zeta\right|,\left|\xi\right| \leq \c n^{-2/3}$ and
$\Im\zeta,\Im\xi \geq 1$.
Then,
\begin{equation}\label{e:FFdiff}
\left|
F_n\BrZW - 4\Fn \BrZetaKsi
\right|\leq  Cn^{-1/6}
\left(
\left|F_n \BrZW\right|
+1
\right).
\end{equation}
\end{prop}

The next step is to prove the convergence of $F_n\BrZW$ to the
transformation $\mathcal{F}$ \eqref{d:KA} of the Airy kernel $\QA$ \eqref{d:AiKer}.
$\mathcal{F}$ can be calculated in terms of the Airy functions, thus
we are concentrated on the calculations of $F_n$.
First, using the properties of CMV matrices,
we present $F_n\BrZW$ in terms of the "resolvent" of $\C$. After that we use
the asymptotic behaviour of the Verblunsky coefficients, obtained
in \cite{Po:12}, to get an approximation of the "resolvent". The approximation
will be given in terms of the Airy functions. Then
we will estimate the error of the "resolvent" approximation and prove
the uniform bound \eqref{e:Fn_conj}.

We start with a simple corollary from the spectral theorem and
Proposition~\ref{p:G_prop}.\smallskip%\medskip

\begin{prop}\label{p:Gn}
Let
\begin{equation*}
\g\left(z\right) = \left(\C + \e{iz}\right)\left(\C - e^{iz}\right)^{-1},
\end{equation*}
be the "resolvent" of the CMV matrix $\C$. Then,
\begin{equation*}
\left(\g \BrZ \right)^{\dagger} =
-\g \left(\overline{z}\right),
\quad
\G \BrZ : = \dfrac{1}{2}
\left(\g \BrZ - \g\left(\overline{z}\right)\right),
\end{equation*}
\begin{equation*}
\g \BrZ \left(\g \BrZ \right)^{\dagger}= -\mathrm{I} + 2\cot \Im z
\cdot \G \BrZ
\end{equation*}
and
\begin{equation}\label{e:FnSum}
F_n\BrZW = n^{-4/3}\sumd\limits_{j,k=0}^{n-1} \G_{j,k}\BrZ \G_{k,j}\BrW.
\end{equation}
\end{prop}

First of all, we would like to restrict the summation above by $j,k \leq M
= \left[C n^{1/2} \log n\right]$ with some constant
$C$.\setcounter{lem}{6}\smallskip

\begin{lem}\label{l:FnSum_est1}
There exists $V$-depended constants $C$ such
that under the conditions of
Theorem \ref{t:main} uniformly in $\Omega_A$ of \eqref{d:On}
we have
\begin{equation*}
n^{-2/3} \sumd_{j=M+1}^{n}
\G_{n-j,n-j}\BrZ \leq Cn^{-1/12} \log n.
\end{equation*}
\end{lem}

Now we present the approximation for the matrix elements $\G_{n-j,n-k}$.
Using the three-diagonal matrices expansion \eqref{e:C_ML} of the $\C$, we
can write the matrix $\g$ as
\begin{equation*}
\g \BrZ = \left(\M \e{-iz/2} + \L \e{iz/2}\right)
\left(\M \e{-iz/2} - \L \e{iz/2}\right)^{-1}.
\end{equation*}
From the definitions of $\M$ and $\L$ one can find their matrix elements
\begin{equation*}
\begin{array}{rl}
\M_{n+k,n+k-1} = \dnk \rhon_{n+k}, & \M_{n+k,n+k}
 = \dnk \alphan_{n+k} - \ndnk \alphan_{n+k+1},
\\
\L_{n+k,n+k-1} = \ndnk \rhon_{n+k}, & \L_{n+k,n+k}
= \ndnk \alphan_{n+k} - \dnk \alphan_{n+k+1},
\end{array}
\end{equation*}
where $d_k = \left(1+s_k\right)/2$ and $s_k=\left(-1\right)^{k}$. Denote
\begin{equation*}
\C_{\pm}\BrZ = \M \e{-iz/2} \pm \L \e{iz/2}.
\end{equation*}

At the first step we derive the representation for the matrix elements of
the inverse matrix of $\C_{-}\BrZ$. Note that $\CRm$ is three-diagonal and
symmetric, and its entries are
\begin{equation*}
\begin{array}{rl}
\C_{-_{n+k,n+k-1}} \BrZ = &
s_{n+k} \rhon_{n+k} e_{n+k}\BrZ,
\\
\C_{-_{n+k,n+k}} \BrZ  = & s_{n+k}\alphan_{n+k}e_{n+k}\BrZ +
s_{n+k}\alphan_{n+k+1}e_{n+k+1}\BrZ
\end{array}
\end{equation*}
with
\begin{equation*}
e_k\BrZ = \cos \dfrac{z}{2} - i s_{k}
\sin \dfrac{z}{2}.
\end{equation*}
For the Verblunsky coefficients we use the result of \cite{Po:12}.\medskip

\begin{lem}\label{l:assym}
Consider the system of orthogonal polynomials and the Verblunsky
coefficients defined above. Let the potential $V$ satisfy conditions {\rm
C1--C3} above. Then, for any $k,$
\begin{eqnarray*}
\alphan_{n+k}&=&\left(-1\right)^{k} \sn
\left(
\cht - \p\xn_k n^{-2/3}
\right)+\Obig\left(\epsnk\right),
\\
\rhon_{n+k}&=&\sht+\ctht\p\xn_kn^{-2/3}
+\Obig\left(\epsnk\right),
\end{eqnarray*}
where $ \sn = 1$ or $ \sn = -1$ and
\begin{equation*}
\xn_{k} = kn^{-1/3},
\quad \epsnk= n^{-4/3} \log^{11}n
\left(1+\left(\xn_k\right)^2\right)
\mathrm{1}_{\left|k\right|< n}+
\mathrm{1}_{\left|k\right|\geq n},
\end{equation*}
with $\p = \dfrac{\pi \sqrt2 }{P\left(\theta\right)}$ and $P$ defined in
\eqref{d:P}.
\end{lem}

To introduce the approximation for the resolvent, we define two "rotation"
matrices which help to present the matrix $\CRm$ in the form, similar to
the discrete Laplacian matrix. Let $\U$ and $\V$ be two semi-infinite
matrices with the entries
\begin{equation*}
\U_{n+j,n+k} = \left(i\sn\right)^{2nk-k-1} \delta_{jk}, \quad
\V_{n+j,n+k} =  \left(i\sn\right)^{2nk-k}\delta_{jk}
\end{equation*}
and
\begin{equation*}
\CRpm \BrZ = \U\C_{\pm}\V, \quad
\Res \BrZeta = \left(
\CRm \left(z\right)
\right)^{-1},
\,
\text{where}
\,
z = \theta + \zeta n^{-2/3}.
\end{equation*}
Then the entries of the new matrix are
\begin{eqnarray*}
\left(\CRm\right)_{n+k,n+k-1}\BrZ &= &
\rhon_{n+k}e_{n+k}\BrZ,
\\
\left(\CRm\right)_{n+k,n+k}\BrZ
& = &
-i\sn s_n
\left(
\alphan_{n+k}e_{n+k}\BrZ
+\alphan_{n+k+1}e_{n+k+1}\BrZ
\right)
.
\end{eqnarray*}
Using the above definitions, we write
\begin{equation}\label{e:gnr}
\g \left(z \right)=
\mathrm{I}+2\L\V\Res\BrZeta\U\e{iz/2}.
\end{equation}

Now we prove that the matrix elements of $\Res \left(\zeta\right)$ can be
expressed in terms of the Airy functions. For this aim we present an
approximation matrix $\Rstar$ and find the difference between $\Rstar$ and
$\Res$. Note that
\begin{eqnarray*}
\e{iz/2} &=&
\e{i\theta/2} + i\e{i\theta/2}\zeta n^{-2/3} +
\Obig\left(\left|\zeta\right|^2 n^{-4/3} \right),
\\
e_{n+k}\BrZ &=&
e_{n+k}\BrTheta - is_{n+k}e_{n+k}\BrTheta\zeta n^{-2/3} +
\Obig\left(\left|\zeta\right|^2 n^{-4/3} \right),
\end{eqnarray*}
Let $\yn_k = \xn_k - n^{-1/3}/2$ and $\rn_{k,\zeta} = n^{-4/3}
\epsnk+\left|\zeta\right|^2$. Then
\begin{eqnarray}
\notag
\left(\CRm\right)_{n-k,n-k-1}\BrZeta  &= &
\sht e_{n+k}\BrTheta
-
\ctht e_{n+k}\BrTheta p_\theta \yn_k n^{-2/3}
\\
\notag
&&- is_{n+k} \sht e_{n+k}\BrTheta
\zeta n^{-2/3}
- \dfrac{1}{2} \ctht e_{n+k}\BrTheta p_\theta n^{-1}\\
\label{e:CRm1}
&&+ n^{-4/3} \Obig \left(\rn_{k,\zeta}\right),
\end{eqnarray}
\begin{align}
&\left(\CRm\right)_{n-k,n-k}\BrZeta = -\st- 2\sht p_\theta \yn_k n^{-2/3}
\notag\\
\label{e:CRm2} &-2\ccht\zeta n^{-2/3} - is_{n+k}p_\theta \cht n^{-1}
+n^{-4/3} \Obig \left(\rn_{k,\zeta}\right).
\end{align}

The matrix elements of $\CRm$ are similar to the matrix elements of the
discrete Laplace operator with some potential in the $n^{-1/3}$ scale, but
off-diagonal elements contain alternating terms $is_{n+k}\ssht$. Hence, we
define the approximate resolvent in terms of the Airy function with some
shift. Set
\begin{equation*}
\dn_k = is_{n+k+1}\delta,\quad \delta=\dfrac{1}{2}\tht,\quad h=n^{-1/3}
\end{equation*}
and
\begin{equation}\label{d:Rstar}
\Rstar_{n-k,n-j}\BrZeta = h^{-1} \Rz \left(\yn_k + \dn_k h,\yn_j + \dn_j h\right),
\end{equation}
where $\Rz \BrZW,$ defined by
\begin{equation}\label{d:Rz}
\Rz\left(z,w\right) = ab^{-1}\pi
\left\{
\begin{array}{ll}
\psi_{-}\left(z,\zeta\right)
\psi_{+}\left(w,\zeta\right),&\, \Re z\leq \Re w,
\\
\psi_{+}\left(z,\zeta\right) \psi_{-}\left(w,\zeta\right),&\, \Re
z \geq \Re w
\end{array}
\right.
\end{equation}
with $\psi_{\pm}$ defined in the Appendix, %\nameref{s:App},
is the extension of the resolvent of the \mbox{operator $\mathcal{L}$}
\begin{equation}\label{d:Lf}
\Lf \left[f\right]\left(x\right) = a^3 f''\left(x\right) - b^3 x
f\left(x\right)
\end{equation}
to the complex plane, where $a^3 = \st$ and $b^3 = 2\p \sin^{-1}
\left(\theta/2\right)$. For the properties, asymptotic behaviour, and the
integral representation of $\Rz$ see~Appendix. %\nameref{s:App}.
 Denote by $\D$ the
error of the approximation
\begin{equation}\label{d:D}
\D \BrZeta = \CRm\BrZeta \Rstar\BrZeta - I.
\end{equation}

To present the bounds for $\D_{n-k,n-j},$ we introduce the notations
\begin{equation*}
d_{n-k,n-j}^{\left(p\right)} =
\sup\limits_{\left|s\right|\leq \delta+1}
\left|
\dfrac{\partial^{p}}{\partial z^p}
\Rz\left(\yn_k+sh,\yn_j+\dn_j h\right)\right|.
\end{equation*}

One can see from the definition of $\Rz$ that $\dfrac{\partial}{\partial z}
\Rz$ is not defined for $z=w$. In this case, by $\dfrac{\partial}{\partial
z}$ we denote the half of the sum of the left and the right derivatives
$\dfrac{1}{2} \left(\dfrac{\partial_{+}}{\partial z}+
\dfrac{\partial_{-}}{\partial z}\right)$. Then $\D$ satisfies the following
bound.\medskip

\begin{lem}\label{l:Dn}
There exists constants $C_1,C_2$ such that uniformly in
$k,j$ and $\zeta \in \Omega_A$
\begin{eqnarray}
\notag
&&\D_{n-k,n-j}\BrZeta \leq C_1 h^2 \log^{C_2} n
\\
\label{e:Dn_est} && \left( \left(1+h^2 \left|\yn_k\right|^2\right)
d_{n-k,n-j}^{\left(0\right)}
+\left(\left|\yn_k\right|+\left|\zeta\right|\right)
d_{n-k,n-j}^{\left(1\right)} \right).
\end{eqnarray}
\end{lem}

Now we are ready to analyse the r.h.s of \eqref{e:FnSum}. From
\eqref{d:Rstar}, \eqref{e:gnr}, and Lemma~\ref{l:Dn} one can see that
$\G_{n-k,n-j} \approx h^{-1} \Im\Rz \left(\yn_k,\yn_j\right)$, and if we could
neglect the remainder, then
\begin{equation*}
F_n \BrZetaKsi \approx h^2 \sumd
\Im\Rz \left(\yn_k,\yn_j\right)
\Im\mathcal{R}_\xi  \left(\yn_j,\yn_k\right).
\end{equation*}

On the other hand, changing a double sum by the double integral and using
\eqref{e:RzInt}, we obtain $\mathcal{F}\left[\QA\right]$. Hence, our main
goal now is to estimate the remainder that appears after replacement of the
"resolvent" of $\CRm$ by the resolvent of the differential operator. We
will do these calculations in several steps.

We start from the proof of the bound for
\begin{equation}\label{e:FnM_b}
\Sigma_M = n^{-2/3} \sumd\limits_{j=0}^{M}
\G_{n-j,n-j} \BrZ
\end{equation}
with $M = \left[C_0n^{1/2}\log n\right]$. It follows from \eqref{e:gnr} and
the definition of $\G$ that
\begin{equation*}
\G \BrZ =
\L\V\left(
\Res\BrZeta \e{iz/2} -
 \Res\left(\overline{\zeta}\right)\e{i\overline{z}/2}
 \right)\U.
\end{equation*}
Using the definition of $\D$, we can write $\Res$ as
\begin{equation*}
\Res \BrZeta = \Rstar - \Res \BrZeta \D \BrZeta.
\end{equation*}
Then,
\begin{equation*}
\Sigma_M = n^{-2/3} \sumd\limits_{j=0}^{M}
\left(\L\V\left(\Rstar_e \BrZeta - \Res_e \D \BrZeta\right)\U\right)_{n-j,n-j}
=\Sigma_M^* - \Sigma_M^{\D},
\end{equation*}
where $\Rstar_e \BrZeta = \Rstar\BrZeta \e{iz/2}-
\Rstar\left(\overline{\zeta}\right) \e{i\overline{z}/2}$ and the same with
$\Res$ and $\Res_e$. Here $\Sigma_M^{\star}$ can be estimated immediately
by using Proposition~\ref{p:Sum_Int}, and $\Sigma_M^{\D}$ can be estimated
by multiplying $\Sigma_M^{1/2}$ by some small factor which we get using the
Cauchy inequality and the bounds \eqref{e:Dn_est} for $\D_{n-k,n-j}$. Thus
we obtain the quadratic inequality \eqref{e:S_sq_in}. Solving this
inequality, we will obtain \eqref{e:FnM_b}. Indeed,
\begin{eqnarray}\label{e:FnM_s_b}
\left|\Sigma^*_M\right| &\leq& C
\sumd\limits_{j=0}^{M} \sumd\limits_{\left|k-j\right|\leq 1}
h \left|\Im \Rz \left(\yn_k+\dn_k h,\yn_j+\dn_j h\right)\right|
\\
&&+ h^3 \left|\Rz \left(\yn_k+\dn_k h,\yn_j+\dn_j h\right)\right|.
\end{eqnarray}
Using Proposition~\ref{p:Sum_Int}, we can estimate $\Sigma_M^{\star}$ as
follows:
\begin{equation*}
\left|\Sigma_M^{*}\right|
\leq C.
\end{equation*}
To estimate $\Sigma_M^{\D}$, we start with the relation
\begin{align*}
\L\V\Res_e\D\U &= \L\V\Res_e\U \left(\U\right)^{-1}\D\U\\ & =
\left(\g\BrZ - \g\left(\overline{z}\right)\right)\widehat{\D},
\end{align*}
where $\widehat{\D}$ entries have the same bounds as $\D,$ and we will
write below $\D$ to simplify notations. Note that
\begin{multline*}
\left(\g\D\right)_{n-j,n-j} = \left\langle\g\D e_{n-j},e_{n-j}\right\rangle
= \left\langle\D e_{n-j},\left(\g\right)^{\dagger} e_{n-j}\right\rangle
\\
\leq
\left\|\D e_{n-j}\right\| \left\|\left(\g\right)^{\dagger} e_{n-j}\right\| =
\left(\left(\D\right)^{\dagger}\D\right)_{n-j,n-j}^{1/2}
\left(\left(\g\right)^{\dagger}\g\right)_{n-j,n-j}^{1/2},
\end{multline*}
and by the Cauchy inequality and \eqref{e:gsq},
\begin{eqnarray*}
\left|\Sigma_M^{\D} \right| &\leq & C n^{-2/3}
\left(
\sumd\limits_{j=0}^{M} \left(\left(\D\right)^{\dagger}\D\right)_{n-j,n-j}
\right)^{1/2}\\
&&\times\left(
M+2\coth \Im z \sumd\limits_{j=M_1+1}^{M_2} \G_{n-j,n-j}
\right)^{1/2}
\\
& = & S_{\D}^{1/2} \left(\Obig\left(n^{-5/6}\log n\right)
+ 2 n^{-2/3} \coth \left(\Im \zeta n^{-2/3}\right) \Sigma_M \right).
\end{eqnarray*}

Using Lemma~\ref{l:Dn}, the Cauchy inequality, and
Proposition~\ref{p:Rz_int}, we estimate $S_{\D}$ as follows:
\begin{eqnarray*}
S_{\D} &=&\sumd\limits_{j=0}^M
\left(\left(\D\right)^{\dagger}\D\right)_{n-j,n-j}
\\
&&
\leq C_1 n^{-4/3} \log^{C_2} n
\sumd\limits_{j=0}^{M}
\sumd\limits_{k=0}^{\infty}
\left(
\left|\yn_k\right|^2+
\left|\zeta\right|^2
\right)\left|d_{n-k,n-j}^{\left(1\right)}\right|^2
+
\left|d_{n-k,n-j}^{\left(0\right)}\right|^2
\\
&&+\,h^4
\left(
\left|\yn_k\right|^4+
\left|\zeta\right|^4
\right)\left|d_{n-k,n-j}^{\left(0\right)}\right|^2
\\
&&
\leq C_1 n^{-1} \log^{C_2} n
\sumd\limits_{j=0}^{M}
\left(
1+\left|\yn_j\right|
\right)^{3/2} +
h^4\left(
1+\left|\yn_j\right|
\right)^{5/2}
\\
&&
 \leq C_1 n^{-2/3} \log^{C_2} n \left(Mn^{-1/3}\right)^{5/2}
 \leq C_1 n^{-1/4} \log^{C_2} n.
\end{eqnarray*}

Combining this inequality with the above estimate of $\Sigma_M^{\D}$, we
obtain the inequality for $\Sigma_M$
\begin{equation}\label{e:S_sq_in}
\left|\Sigma_M\right| \leq C_1 + C_2 n^{-1/8} \log^{C_3} n
\left(\Obig\left(n^{-5/6}\log
n\right)+\left|\Sigma_M\right|\right)^{1/2}
\end{equation}
which gives \eqref{e:FnM_b}.

Now we are ready to find the limit of the r.h.s. of \eqref{e:FnSum}.
Combining Lemma~\ref{l:FnSum_est1}
with \eqref{e:FnM_s_b}, we get
\begin{equation}\label{e:Gjj}
n^{-2/3}\sumd\limits_{j=0}^{n} \G_{n-j,n-j}\BrZ \leq C.
\end{equation}

Using the definition of $\G$, the sum in \eqref{e:FnSum} can be splitted
into four parts with different products of $\g$ and $\overline{\g}$. For
each sum, the Cauchy inequality yields
\begin{eqnarray*}
&&n^{-4/3}\left|
\sumd\limits_{j,k} \g_{n-j,n-k}\BrZ \g_{n-k,n-j}\BrW
\right|
\\
&&\leq  \left(n^{-4/3} \sumd\limits_{j}
\left(\g\left(\g\right)^{\dagger}\right)_{n-j,n-j}\BrZ \right)^{1/2}\\
&&\times\left(n^{-4/3} \sumd\limits_{j}
\left(\g\left(\g\right)^{\dagger}\right)_{n-j,n-j}\BrW \right)^{1/2}
,
\end{eqnarray*}
where each of the brackets is bounded because of \eqref{e:gsq} and
\eqref{e:Gjj}. Changing the summation limits in the previous bound to $j
\in [M,n]$ and using Lemma~\ref{l:FnSum_est1}, we obtain that under the
conditions of Lemma~\ref{l:ker}
\begin{equation*}
F_n\BrZW = n^{-4/3}\sumd\limits_{j,k=0}^{M} \G_{n-k,n-j}\BrZ \G_{n-j,n-k}\BrW + \Obig\left(n^{-1/24}\log n\right).
\end{equation*}

Now we use once more the identity
\begin{equation*}
\G = G^{\star} - \G\widehat{\D}.
\end{equation*}
Repeating the above arguments, we obtain
\begin{equation*}
F_n\BrZW = F_n^{\star}\BrZW + F_{\D}\BrZW,
\end{equation*}
and
\begin{equation*}
 F_{\D}\BrZW \leq C_1 n^{-1/8} \log^{C_2} n.
\end{equation*}
Since $G^{\star} = \L\V\Rstar_e \U$ with $\Rstar_e$ defined above, we have
\begin{equation*}
G^{\star}_{n-k,n-j} = n^{1/3} \Im \Rz\left(\yn_k,\yn_j\right) +
r^{G^{\star}}_{k,j},
\end{equation*}
where $r^{G^{\star}}_{k,j}$ contains terms with some derivatives of the
$\Rz$ multiplied by $h$ in some non-negative power. Thus, from the
boundness of the corresponded integrals (see proof of
Proposition~\ref{p:Rz_int} for the arguments)
\begin{equation*}
h^{p+q} \intd\limits_{0}^{Mn^{-1/3}}\intd\limits_{0}^{Mn^{-1/3}}
\left|\dfrac{\partial^{p+q}}{\partial x^p \partial y^q} \Rz\BrXY\right|^2
 dx dy \leq C_{p,q,r,s},
\end{equation*}
we obtain that we can neglect terms from $r^{G^{*}}_{k,j}$ and
\begin{equation*}
F_n^{\star} \BrZW = \intd\limits_{0}^{Mn^{-1/3}}\intd\limits_{0}^{Mn^{-1/3}}
 \Im \Rz \BrXY
\Im \mathcal{R}_{\xi} \left(y,x\right) dxdy + \Obig\left(h^{1/2}\right).
\end{equation*}

Finally we note that by \eqref{e:prod} and \eqref{e:Rxx},
\begin{equation*}
\intd\limits_{Mn^{-1/3}}^{\infty} dx \intd dy
\left|\Rz \BrXY\right|^2
\leq
\intd\limits_{Mn^{-1/3}}^{\infty}
\Im \Rz \left(x,x\right) dx \leq C n^{-1/12} \log n,
\end{equation*}
and
\begin{equation*}
\intd\limits_{0}^{\infty} \intd\limits_{0}^{\infty}
\Im \mathcal{R}_{\zeta} \left(x,y\right)
\Im \mathcal{R}_{\xi} \left(y,x\right)
dx dy \leq C.
\end{equation*}
Hence,
\begin{equation}\label{e:FnF}
F_n \BrZW = \intd\limits_{0}^{\infty} \intd\limits_{0}^{\infty}
\Im \mathcal{R}_{\zeta} \left(x,y\right)
\Im \mathcal{R}_{\xi} \left(y,x\right)
dx dy + \Obig\left(C n^{-1/24}\log^C n\right).
\end{equation}
Estimate \eqref{e:FnF}, integral representation \eqref{e:RzInt}, and the
following relation (see \cite{TW:94})
\begin{equation*}
\QA \left(x,y\right) = \intd\limits_{0}^{\infty}
Ai \left(x+ t \right)
Ai \left(y+ t \right) dt
\end{equation*}
imply \eqref{e:Fn_conj} with
\begin{equation*}
\mathcal{K} \BrXY = a^{-2}b^{-4}
\QA \left(a^{-1}b^{-2}x,a^{-1}b^{-2}y\right).
\end{equation*}

Proposition~\ref{p:Fred} implies that it is sufficient to check
\eqref{e:Fr_conj} to finish the proof of Theorem~\ref{t:main}. We use an
evident relation
\begin{equation*}
G\left(t+i\varepsilon - s\right) = \dfrac{d}{dt}
2\arctan\left(\tan\left(\dfrac{t-s}{2}
\right)\cot\dfrac{\varepsilon}{2}\right)
\end{equation*}
that implies the inequality valid for any $s \in \left[a,b\right] \subset
\mathbb{R}$
\begin{equation*}
\intd\limits_{a-1}^{b+1} G \left(\left(t+i-s\right)n^{-2/3}\right)dt
\geq C n^{2/3},
\end{equation*}
with some absolute constant $C$. The last inequality, the positiveness of
$\mathcal{K}_n$ and $G$, and definition of $\G$ imply
\begin{eqnarray*}
\intd\limits_{a}^{b}
\mathcal{K}_n \left(s,s\right) ds
&&\leq C n^{-2/3} \intd\limits_{a}^{b} ds \intd\limits_{a-1}^{b+1} dt
\mathcal{K}_n\left(s,s\right)G \left(\left(t+i-s\right)n^{-2/3}\right)
\\
&&
\leq
C \intd\limits_{a-1}^{b+1} \sumd\limits_{j=1}^{n}
\G_{n-j,n-j} \left(\theta+\left(t+i\right)n^{-2/3}\right) dt.
\end{eqnarray*}
Hence, by \eqref{e:Gjj} for any finite $\Delta \subset
\left[-A+1,A-1\right]$ we obtain
\eqref{e:Fr_conj}.\smallskip\setcounter{equation}{0}

{\centering\subsection{Auxiliary Results}\label{s:aux}}

P r o o f of Proposition~\ref{p:G_est}. Using Lemma~\ref{l:Kn_int} with
$\varepsilon = 2\c$ and inequality
\begin{equation}\label{e:Kn_ci}
\left|\K_n\BrLambdaMu\right|^2 \leq \K_n\left(\lambda,\lambda\right)
\K_n\left(\mu,\mu\right),
\end{equation}
we obtain
\begin{equation*}
\intd\limits_{\lambda \in \sigma_{\varepsilon}^{c}}
G\left(z-\lambda\right)
\left|\K_n\BrLambdaMu\right|^2 d\lambda
\leq C \e{-nd\left(\varepsilon\right)}
\sup\limits_{\lambda \in \sigma_{\varepsilon}^{c}}
G\left(z-\lambda\right)
\K_n \left(\mu,\mu\right).
\end{equation*}
Due to the restrictions on $\lambda$ and $z$ we get
$G\left(z-\lambda\right)\leq C'$ when $\lambda \in \sigma_{\varepsilon}^c$.
Thus,
\begin{equation*}
\iintd\limits_{\sigma_{\varepsilon}^{c}}
G\left(z-\lambda\right)
G\left(w-\mu\right)
\left|\K_n \BrLambdaMu\right|^2 d\lambda d\mu
=\e{-cn}\Obig\left(\Im^{-1}z+\Im^{-1}w\right).
\end{equation*}
Changing the variables by the scaled ones in \eqref{d:Fn2}, we get
\begin{equation*}
F_n\BrZW =
n^{-4/3} \iintd\limits
\Im \cot\dfrac{\zeta-x}{2n^{2/3}}
\Im \cot\dfrac{\xi-y}{2n^{2/3}}
\left|\Kn\BrXY\right|^2 dxdy+\Obig\left(\e{-cn}\right).
\end{equation*}

Finally we estimate the difference between $F_n$ and $4\Fn$
\begin{equation*}
4\Fn\BrZetaKsi - F_n\BrZW  = n^{-4/3}\left(I_1\BrZetaKsi
+I_2\BrZetaKsi+I_2\left(\xi,\zeta\right)\right)
+\Obig\left(\e{-cn}\right)
\end{equation*}
with $I_1$ and $I_2$ of \eqref{e:I_1} and \eqref{e:I_2}. It is easy to see
that
\begin{multline}\label{e:I_1}
\left|I_1\BrZetaKsi\right| \!=\! \left| \iintd\!\! \Im \!\left(
\dfrac{2n^{2/3}}{\zeta-x}- \cot\dfrac{\zeta-x}{2n^{2/3}} \right)\! \!\Im \!
\left( \dfrac{2n^{2/3}}{\xi-y}- \cot\dfrac{\xi-y}{2n^{2/3}} \right)\!
\left|\Kn\BrXY\right|^2 dxdy \right|
\\
\leq C \iintd\limits
\left|\Kn\BrXY\right|^2 dxdy
\leq Cn,
\end{multline}
where we have used that for $0<\left|z\right|\leq 2\c$
\begin{equation*}
\left|\cot z - \dfrac{1}{z}\right| \leq C.
\end{equation*}

In addition, since the kernel $\left|\K_n\BrLambdaMu\right|^2$ is positive
definite, we can use the Cauchy inequality to get
\begin{multline}\label{e:I_2}
\left|I_{2}\BrZetaKsi \right|=
\left|
\iintd
\Im \left(
\dfrac{2n^{2/3}}{\zeta-x}-
\cot\dfrac{\zeta-x}{2n^{2/3}}
\right)
\Im
\cot\dfrac{\xi-y}{2n^{2/3}}
\left|\Kn\BrXY\right|^2 dxdy
\right|
\\
\leq \left|I_1\BrZetaKsi\right|^{1/2} \left|n^{4/3} F_n\BrZW\right|^{1/2}
\leq Cn^{7/6} \left|F_n\BrZW\right|^{1/2}.
\end{multline}

Finally, collecting the above bounds, we obtain
\begin{equation*}
\left|F_n\BrZW - \Fn\BrZetaKsi\right|
\leq Cn^{-1/6} \left|F_n\BrZW\right|^{1/2}
+C'n^{-1/3},
\end{equation*}
and using the Cauchy inequality, we get
\eqref{e:FFdiff}.\hfill\rule{0.5em}{0.5em}\newpage%\smallskip

P r o o f of Lemma~\ref{l:Dn}. The proof is based on the direct
calculations of the matrix elements $\D_{n-j,n-k}$. We start with the case
$j \neq k$. Then all derivatives of $\Rz$ are well defined and the points
$\yn_{j-1},\yn_j, \yn_{j+1}$ are laying on the same side of $\yn_k$. Now we
are going to calculate $\D_{n-j,n-k}$ using the Taylor expansion and
definition of the $\CRm$. These calculations are a little bit involved, so
we present them in several steps. First, we calculate $\Rstar_{n-k\mp 1,
n-j}$,
\begin{gather*}
\Rstar_{n-k\mp 1,n-j} = h^{-1} \Rz \left(\yn_k \pm h - \dn_k h,\yn_j+\dn_j
h\right)
\\
= h^{-1} \Rz \left(\yn_k,\yn_j+\dn_j h\right) + \left(\pm 1 - \dn_k\right)
\dfrac{\partial}{\partial z}\Rz\left(\yn_k,\yn_j+\dn_jh\right)
\\
+\left(\pm 1 - \dn_k\right)^2 h \dfrac{\partial^2}{\partial
z^2}\Rz\left(\yn_k,\yn_j+\dn_jh\right) + h^2\Obig\left(\rstar_{n-k,n-j}
\left(\delta+1\right)\right)
\end{gather*}
with the remainder
\begin{equation*}
\rstar_{n-k,n-j}\left(d\right) = \sup\limits_{\left|s\right|<d}
\left|\dfrac{\partial^3}{\partial z^3}
\Rz\left(\yn_k+s,\yn_j+\dn_jh\right)\right|,
\end{equation*}
where the last bound follows from  differential equation \eqref{e:Aeq}
valid for the functions $\psi_{\pm}$. To simplify calculations for $\CRm$,
we use the following notations:
\begin{gather*}
S_k:= \left(\CRm\right)_{n-k,n-k-1}+ \left(\CRm\right)_{n-k,n-k+1}, \\
D_k:= \left(\CRm\right)_{n-k,n-k-1}- \left(\CRm\right)_{n-k,n-k+1}.
\end{gather*}

Then, combining the above expansion with \eqref{e:CRm1}--\eqref{e:CRm2}, we
obtain
\begin{align}%{eqnarray}
\notag
&\D_{n-k,n-j} = h^{-1} \Rz \left(\yn_k,\yn_j+\dn_j h\right)
\left(S_k+\left(\CRm\right)_{n-k,n-k}  \right)
\\
\notag
&+\dfrac{\partial}{\partial z}\Rz \left(\yn_k,\yn_j+\dn_j h\right) \left(
D_k - \dn_k S_k +\dn_k \left(\CRm\right)_{n-k,n-k} \right)
\\
\notag
&+h\dfrac{\partial^2}{\partial z^2}\Rz \left(\yn_k,\yn_j+\dn_j h\right)
\left(
\dfrac{1}{2}S_k - \dn_k D_k -
\dfrac{\delta^2}{2}\left(S_k+\left(\CRm\right)_{n-k,n-k}\right)
\right)
\\
\label{e:dkj}
&+ \Obig\left(\rstar_{n-k,n-j}\left(\delta+1\right)\right),
\end{align}%{eqnarray}
where for the last term we have used the uniform bound for elements
$\bigl(\CRm\bigr)_{n-j,n-k}$.

Now it is sufficient to calculate every expression in the brackets. We
start with $S_k$ and $D_k$,
\begin{equation*}
S_k=\st - 2\cht\ctht \p \yn_k h^2 -
2\ssht \zeta  h^2 + is_{n+k}\p \cht h^3 +
h^4\Obig\left(\rn_{k,\zeta}\right),
\end{equation*}
\begin{gather*}
D_k=-2is_{n+k}\ssht + 2is_{n+k}\cht \p \yn_k h^2 -is_{n+k} \st \zeta h^2
\\ -\cht\ctht\p h^3+ h^4\Obig\left(\rn_{k,\zeta}\right).
\end{gather*}

Therefore, with an error of order $h^4 \Obig\left(\rn_{k,\zeta}\right)$ we
can write
\begin{equation*}
S_k + \left(\CRm\right)_{n-k,n-k} \approx -2 h^2
\left(
\p \sin^{-1} \left(\theta/2\right)\yn_k + \zeta
\right),
\end{equation*}
\begin{eqnarray*}
D_k - \dn_k S_k + \dn_k \left(\CRm\right)_{n-k,n-k} &\approx&
-2\dn_k h^2 \left(\p \sin^{-1} \left(\theta/2\right) \yn_k - \zeta
\right.
\\
&&
\left.
+\,
is_{n+k} \p \cos \left(\theta/2\right) \sin^{-2} \left(\theta/2\right) h\right) .
\end{eqnarray*}

Finally, combining the above relations and the equation for $\Rz$ in the
form
\begin{gather*}
\st \dfrac{\partial^2}{\partial z^2}\Rz \left(\yn_k,\yn_j+\dn_j h\right)\\
-\left(2\p \sin^{-1} \theta/2 \yn_k + \zeta \right) \Rz
\left(\yn_k,\yn_j+\dn_j h\right)=0,
\end{gather*}
we obtain the remainder in \eqref{e:dkj} with all terms of order less than
$h^2$. Gathering all these remainders and the remainder $h^4
\Obig\left(\rn_{k,\zeta}\right),$ we get \eqref{e:Dn_est}. For $j=k$, the
calculations can be performed similarly if we take into account jump
condition \eqref{e:Ajump}.\hfill\rule{0.5em}{0.5em}\smallskip

P r o o f of Lemma~\ref{l:FnSum_est1}. We start with estimate of
\begin{equation*}
X_n \BrZeta = n^{-2/3} \intd \Kn\left(x,x\right)
G\left(\left(\zeta-x\right)n^{-2/3}\right) dx,
\end{equation*}
where $\Kn$ is defined as in \eqref{d:Kn} but without any restriction. Let
$\zeta = s + i\varepsilon$. Changing variables to $z = \theta + \zeta
n^{-2/3}$ and using \eqref{d:Kn} with \eqref{d:G}, we obtain
\begin{equation*}
X_n\BrZeta = n^{1/3} \Re h_n\BrZ,
\end{equation*}
where
\begin{equation*}
h_n \left( z \right) = \int\limits_{-\pi}^{\pi}
g\left(z-\lambda\right) \, \rho_n \left( \lambda \right) \, d
\lambda.
\end{equation*}

For further estimates we use the "quadratic" equation obtained in
\cite{Po:08},
\begin{equation*}
h^2_n \BrZ -
2iV'\left( \Re z \right)h_n \BrZ-2iQ_n \BrZ -1
=-\dfrac{2}{n^2} \delta_n\BrZ,
\end{equation*}
with
\begin{equation*}
Q_n \BrZ =
\intd\limits_{-\pi}^{\pi}
g\left(z -\lambda \right)
\left(
V' \left( \lambda \right) -
V' \left ( \Re z \right)\right)
 \, \rho_n
\left( \lambda \right) \, d \lambda,
\end{equation*}
\begin{equation*}
\delta_n \BrZ =
\iintd\limits_{-\pi}^{\pi}
\left| \K_n\left( \lambda, \mu \right ) \right|^2
\left(g\left(z-\lambda\right) - g\left(z-\mu\right)\right)^2
\, d \lambda d \mu .
\end{equation*}

Solving the "quadratic" equation, we get
\begin{equation*}
X_n\BrZeta = n^{1/3} \Re
\sqrt{f_n\left(s,\varepsilon\right)- 2n^{-2} \delta_n \BrZ},
\end{equation*}
where the function
\begin{equation*}
f_n\left(s,\varepsilon\right) = -V'^2\left(\theta +
sn^{-2/3}\right)+
2iQ_n\left(\theta+\left(s+i\varepsilon\right)n^{-2/3}\right) + 1
\end{equation*}
is twice differentiable in both variables. Using the symmetry of the kernel
$\K_n$ and \eqref{e:Kn_ci}, we can estimate $\delta_n\BrZ$ as
\begin{equation*}
\left|n^{-2}\delta_n\BrZ\right|
\leq
4 n^{-2} \intd\limits_{-\pi}^{\pi}
\K_n\left(\lambda,\lambda\right)
\left| g\left(z-\lambda\right)
\right|^2
d\lambda.
\end{equation*}

Then the identity \eqref{e:gsq} yields
\begin{equation*}
\left|n^{-2}\delta_n\BrZ\right| \leq 4n^{-1}
+  2n^{-4/3} \coth \left(\varepsilon n^{-2/3} \right)
 \cdot X_n\BrZeta
 \leq Cn^{-2/3}\!\left(\!n^{-1/3}+\varepsilon^{-1} X_n\BrZeta\!\right)\!,
\end{equation*}
as $\varepsilon = \Obig\left(1\right)$. Now we continue the estimation of
$Q_n\BrZ$. For the density $\rho_n$, we use the bound (see \cite{Po:08})
\begin{equation*}
\left|\rho_n'\BrLambda\right| \leq
C\left(
\left|\psi_{n-1}^{\left(n\right)}\right|^2
+\left|\psi_{n}^{\left(n\right)}\right|^2
+1
\right),
\end{equation*}
where $\psi_{k}^{\left(n\right)}=\P_k w_n^{1/2}$ are orthonormal functions.
Hence, the density $\rho_n$ is uniformly bounded and therefore, similarly
to $\left(2.17\right)$ of \cite{Po:08}, we have
\begin{equation*}
\left|Q_n \left(z\right) - Q_n \left(\Re z\right)\right|
\leq C \Im z \left| \log \Im z \right|.
\end{equation*}

The weak convergence \eqref{e:wconv} with
$$ \phi\BrLambda =
\left(V'\BrLambda - V'\left(\theta+s/\gamma n^{2/3}\right)\right) \cot
\dfrac{\lambda - \theta - s/\gamma n^{2/3}}{2}
$$ implies
\begin{equation*}
\left| Q_n\left(\theta+s/\gamma
n^{2/3}\right)-Q\left(\theta+s/\gamma n^{2/3}\right) \right| \leq
C n^{-1/2} \log^{1/2} n
\end{equation*}
if $\left|s\right| \leq c_\theta n^{2/3}$. Hence, combining the above
relations, we obtain
\begin{equation*}
\left|f_n\left(s,\varepsilon\right) - f\left(s\right)\right|
\leq C n^{-2/3}\log n
\left(
\left|\log\varepsilon\right| + n^{1/6}
\right),
\end{equation*}
with $f\left(s\right):=f\left(s,0\right)$. The properties of the Herglotz
transformation yield (see \cite{Po:08})
\begin{equation*}
\rho\left(\lambda\right) = \dfrac{1}{2\pi}
\lim\limits_{\varepsilon \to +0}\Re
h\left(\lambda+i\varepsilon\right).
\end{equation*}

Therefore, at the edge point $\theta$ we obtain $f\left(0\right)=0$ and
$f'\left(0\right) < 0$. Hence, by the differentiability of
$f\left(s\right),$ we obtain
\begin{equation}\label{e:Xsqrt}
X\BrZeta = \Re\sqrt{
\Obig\left(s+\varepsilon^{-1} X\BrZeta +
n^{1/6} \log n
\right)}.
\end{equation}

Solving the quadratic inequality, we estimate $X\BrZeta$ as follows:
\begin{equation*}
X\BrZeta \leq C \left(
\varepsilon^{-1} +
s^{1/2}+n^{1/12}\log^{1/2}n
\right).
\end{equation*}

Now we write \eqref{e:Xsqrt} more precisely
\begin{equation*}
X\BrZeta =
\Re\sqrt{-Cs+\varepsilon^{-2}
\Obig\left(1+\varepsilon s^{1/2} +
\varepsilon n^{1/12} \log^{1/2}n \right)}.
\end{equation*}

Below we need the estimate of $X\BrZeta$ for $s>C n^{1/6} \log n$ and
$\varepsilon = \Obig\left(1\right)$. Hence we obtain
\begin{equation}\label{e:Xest}
X\BrZeta \leq C_1 \left|s-C_2n^{1/6}\log n\right|^{-1/2}.
\end{equation}

Note that all constants in the above estimates depend only on $V$ and can
be bounded by some combination of $\sup\left| V \right|$, $\sup \left| V''
\right|$ and $\sup \left| V''' \right|$. Now we return to the estimate of
the sum in Lemma~\ref{l:FnSum_est1}. By the spectral theorem,
\begin{equation*}
I\left(M\right) = n^{-2/3} \sumd_{j=M+1}^{n}
\G_{n-j,n-j}\BrZ
=n^{-2/3}\sumd_{j=0}^{n-M-1}
\intd G\left(\lambda-z\right)
\left|\chi_{j}^{\left(n\right)}\BrLambda\right|^2
w_n\BrLambda d\lambda.
\end{equation*}

Let us consider the analogue of the joint eigenvalue distribution of  model
\eqref{d:Model} in the form
\begin{equation*}
p_{n-M}^{\left(n-M\right)} \left( \lambda_1, \ldots ,
\lambda_{n-M}\right)\!=\!\dfrac{1}{Z_n^{\left(n-M\right)}}
\!\!\prodd\limits_{1 \leq j < k \leq n-M}\!
\left| e^{i \lambda_j} -  e^{i \lambda_k}\right|^2
\!\exp\!
\left\lbrace
\!\! -n \!\!\sum\limits_{j=1}^{n-M}\! V \! \left( \cos\lambda_j
\!\right)
\right\rbrace.
\end{equation*}

Then, by the same argument as above for  model \eqref{d:Model}, we define
the first marginal density
\begin{equation*}
\rho^{\left(n-M\right)}_{n-M}\BrLambda =
\dfrac{1}{n-M} \sumd\limits_{j=0}^{n-M-1}
\left|\chi_{j}^{\left(n\right)}\BrLambda\right|^2
w_n\BrLambda.
\end{equation*}

On the other hand, this density can be considered as the first marginal
density for  model \eqref{d:Model} with the potential
$\widetilde{V}=\dfrac{n}{n-M}V$. Hence,
\begin{equation*}
I\left(M\right) = n^{-2/3} \intd G\left(\lambda-z\right)
K^{\left(n-M,\widetilde{V}\right)}_{n-M} \left(\lambda,\lambda\right)
d\lambda = X_{n-M}^{\widetilde{V}} \BrZeta.
\end{equation*}
But it follows from the result of \cite{KM:00} that the support of the
equilibrium density for $\widetilde{V}$ is $\left[\theta_M,\theta_M\right]$
with $\theta_M = \theta - c_V \left(Mn^{-1}\right)+\Osmall
\left(Mn^{-1}\right)$ with some $c_V
>0$. Hence, by \eqref{e:Xest},
\begin{equation*}
X_{n-M}^{\widetilde{V}}\leq Cn^{-1/12},
\end{equation*}
and Lemma~\ref{l:FnSum_est1} is proved.\hfill\rule{0.5em}{0.5em}\smallskip
\setcounter{equation}{0}\setcounter{prop}{0}\smallskip

{\centering\subsection{Appendix}\label{s:App}}

In this section we present the properties and the asymptotic analysis of
the resolvent of the Airy operator. Denote by $\Lf$ the second order
differential operator on the set of twice continuously differentiable
functions on $\mathbb{R},$
\begin{equation*}
\Lf \left[f\right]\left(x\right)
= a^3 f''\left(x\right) -
b^3 x f\left(x\right).
\end{equation*}

Let $\Rz\BrXY$ be the kernel of the resolvent $\left(\Lf-\zeta
I\right)^{-1}$ for $\Im \zeta \neq 0$. By the general principles (for
example see \cite{Zw:97}, Section 72)\medskip

\begin{prop}\label{p:Rairy}
Let $Ai\left(z\right)$ and $Bi\left(z\right)$ be the standard Airy
functions. Denote by $\psi_{\pm}$ the following functions:
\begin{equation*}
\psi_{-} \left(x,\zeta\right) =
Ci \left(X_{x,\zeta}\right),
\quad \psi_{+}
\left(x,\zeta\right) =
Ai \left(X_{x,\zeta}\right),
\end{equation*}
with
\begin{equation*}
Ci\left(X\right) = iAi\left(X\right) -
Bi\left(X\right)
\quad
\text{and} \quad
X_{x,\zeta} = a^{-1} b x + a^{-1} b^{-2} \zeta.
\end{equation*}
Then these functions are the unique solutions of the differential equation
\begin{equation}\label{e:Aeq}
a^3 \dfrac{\partial^2}{\partial x^2}
\psi_{\pm} \left(x,\zeta\right)
- \left(b^3 x + \zeta\right)\psi_{\pm}
\left(x,\zeta\right) = 0,
\end{equation}
that are square integrable on the right (left) half axis and fixed by jump
condition
\begin{equation}\label{e:Ajump}
\psi_{-}\left(x,\zeta\right)
\dfrac{d}{dx} \psi_{+}\left(x,\zeta\right)-
\psi_{+}\left(x,\zeta\right)
\dfrac{d}{dx} \psi_{-}\left(x,\zeta\right) =
a^{-1} b \pi^{-1}.
\end{equation}
And the resolvent $\Rz$ has two representations
\begin{equation}\label{d:Rez}
\Rz\BrXY = ab^{-1}\pi
\left\{
\begin{array}{ll}
\psi_{-}\left(x,\zeta\right)
\psi_{+}\left(y,\zeta\right),&\, x\leq y,
\\
\psi_{+}\left(x,\zeta\right) \psi_{-}\left(y,\zeta\right),&\,
x\geq y,
\end{array}
\right.
\end{equation}
\begin{equation}\label{e:RzInt}
\Rz \BrXY = a^{-2}b^{-1} \intd \dfrac{1}{t-\zeta} Ai
\left(a^{-1}bx+a^{-1}b^{-2} t \right) Ai
\left(a^{-1}by+a^{-1}b^{-2} t \right) dt.
\end{equation}
\end{prop}

The following asymptotic behaviour of the Airy functions can be found
\mbox{in \cite{Ab-St:65}}.\medskip

\begin{prop}\label{p:AiryAsymp}
For any $\delta>0,$ the following asymptotics are uniform in the
corresponding domains:
\begin{eqnarray*}
Ai\BrZ =& \pi^{-1/2}z^{-1/4}\e{-\frac{2}{3}z^{3/2}}
\left(
1+\Obig\left(z^{-3/2}\right)
\right), &\left|arg z\right|<\pi-\delta,
\\
Ai\left(-z\right) =& \pi^{-1/2}z^{-1/4}
\sin\left(\dfrac{2}{3}z^{3/2}+\dfrac{\pi}{4}\right)
\left(
1+\Obig\left(z^{-3/2}\right)
\right),& \left|arg z\right|<\dfrac{2}{3}\pi-\delta,
\\
Ci\BrZ =& \pi^{-1/2}z^{-1/4}\e{\frac{2}{3}z^{3/2}}
\left(
1+\Obig\left(z^{-3/2}\right)
\right),& \left|arg z\right|<\dfrac{1}{3}\pi-\delta,
\\
Ci\left(-z\right) =& \pi^{-1/2}z^{-1/4}
\e{i\frac{2}{3}z^{3/2}+i\dfrac{\pi}{4}}
\left(
1+\Obig\left(z^{-3/2}\right)
\right),& \left|arg z\right|<\dfrac{2}{3}\pi-\delta.
\end{eqnarray*}
\end{prop}

The main term for the derivatives can be obtained by direct differentiation
of the asymptotics. The last proposition and the definition of the
functions $\psi_{\pm}$ yield the asymptotic behaviour of them\medskip

\begin{prop}\label{p:Psi_pm}
The functions $\psi_{\pm}$ are entire in $x$ and $\zeta$ and have
the following asymptotic behaviour in x for $\Im \zeta =
\varepsilon >0$:
\begin{gather*}%{eqnarray*}
\left|\psi_{+} \left(x,\zeta\right)\right| =\\
\pi^{-1/2}\left|X_{x,\zeta}\right|^{-1/4} \left(
1+\Obig\left(\left|X_{x,\zeta}\right|^{-3/2}\right) \right) \left\lbrace
\begin{array}{ll}
\exp\left\{-\dfrac{2}{3}\left|\Re X_{x,\zeta}\right|^{3/2}\right\}, & x \to \infty \\[11pt]
\exp\left\{a^{-1} b^{-2} \varepsilon \left|\Re X_{x,\zeta}\right|^{1/2}\right\}, & x \to -\infty
\end{array}
\right.
\\
\left|\psi_{-} \left(x,\zeta\right)\right|=\\
\left(4\pi\right)^{-1/2}\left|X_{x,\zeta}\right|^{-1/4}\! \left(\!
1+\Obig\left(\left|X_{x,\zeta}\right|^{-3/2}\right)\! \right)\!
\left\lbrace\!\!
\begin{array}{ll}
\exp\left\{\dfrac{2}{3}\left|\Re X_{x,\zeta}\right|^{3/2}\right\}, & x \to \infty \\ [11pt]
\exp\left\{-a^{-1} b^{-2} \varepsilon \left|\Re X_{x,\zeta}\right|^{1/2}\right\}, & x \to -\infty
\end{array}
\right.
\end{gather*}%{eqnarray*}
\end{prop}

\begin{prop}\label{p:Rz_int}
For any non-negative integers $s,q$ and any $A \in \mathbb{R_{+}}$
there exists a constant $C_{A,s,q}$ such that for any $x\geq -A$
and $\zeta \in \Omega_A$
\begin{equation}\label{e:Rz_int}
I\left(s;q\right) = \intd\limits_{-\infty}^{\infty}
\left|y\right|^s \left| \dfrac{\partial^{q}}{\partial y^q} \Rz
\left(x,y\right)\right|^2 dy \leq C_{A,s,q} \left(
1+\left|x\right|\right)^{s+q-3/2}.
\end{equation}
\end{prop}

P r o o f of Proposition~\ref{p:Rz_int}. In view of  equation
\eqref{e:Aeq}, two extra derivatives in \eqref{e:RzInt} give the extra
factor of order $\left|y\right|^2+\left|\zeta\right|^2$ to the integrand.
Therefore, we start with $I\left(s;0\right)$. Since $\left|\Rz
\left(x,y\right)\right| \leq C_A \e{-c_A \left|x-y\right|^{1/2}}$ for $x
\geq -A$ and $\zeta \in \Omega_A$, we split the integral from
\eqref{e:Rz_int} into two parts
\begin{eqnarray}
\notag
I\left(s;0\right) &= &
\intd\limits_{\left|y-x\right|<2\left|x\right|}
+ \intd\limits_{\left|y-x\right|>2\left|x\right|}
\leq C_s \left(x^s +\left|\zeta\right|^s\right)
\intd\left|\Rz \left(x,y\right)\right|^2 dy
\\
\label{e:RInt_s}
&&+ \, C_A\intd\limits_{t>2\left|x\right|}
\left(t+x\right)^s \e{-c_A t^{1/2}} dt.
\end{eqnarray}

For the first integral we note that by the spectral theorem and the
resolvent identity,
\begin{equation}\label{e:prod}
\intd\limits_{-\infty}^{\infty}\left|\Rz \left(x,y\right)\right|^2 dy
= \dfrac{\Im \Rz \left(x,x\right)}{\Im \zeta}.
\end{equation}

The asymptotic behaviour of $\psi_{\pm}$ from Proposition~\ref{p:Psi_pm}
implies
\begin{equation}\label{e:Rxx}
\left|\Rz \left(x,x\right)\right|\leq
C_A
\left( 1+\left|x\right|\right)^{-1/2},
\quad
\text{and}
\quad
\left|\Im \Rz \left(x,x\right)\right|\leq
C _A
\left( 1+\left|x\right|\right)^{-3/2}.
\end{equation}

Combining \eqref{e:RInt_s} with \eqref{e:prod} and \eqref{e:Rxx}, we obtain
\eqref{e:Rz_int} with $q=0$. In view of equation \eqref{e:Aeq}, it is
sufficient to prove \eqref{e:RzInt} only for $q = 0,1$. If $q=1$, similarly
to the above argument, we split the integral into two parts. In the first
term, integrating by parts, we have
\begin{equation*}
\intd\limits_{-\infty}^{\infty}
\left|\dfrac{\partial}{\partial y}\Rz \left(x,y\right)\right|^2 dy
= \intd\limits_{-\infty}^{\infty}
 \left(c_1 y + c_2\zeta\right)\left|\Rz \BrXY\right|^2 dy.
\end{equation*}

The r.h.s satisfies the necessary bound for $q=1$, hence the proposition is
proved. $~~~~$\hfill\rule{0.5em}{0.5em}\medskip

\begin{prop}\label{p:Sum_Int}
Let $h=n^{-1/3}$, $M = \left[C_0 n^{1/2} \log n\right]$. Also,
denote by $x_j = j h$ the equidistant set and
$z^{\left(1,2\right)}_j = x_j + \delta^{\left(1,2\right)}_j h$ two
shifted sets, with complex shifts
$\left|\delta^{\left(1,2\right)}_j\right|\leq C$ for some absolute
constant $C$. Then,
\begin{equation}\label{e:ImRzz}
h \sumd\limits_{j=0}^{M}
\left|\Im \Rz \left(z^{\left(1\right)}_j,z^{\left(2\right)}_j\right)\right|
\leq C,
\end{equation}
\begin{equation}\label{e:Rzz}
h \sumd\limits_{j=0}^{M}
\left|\Rz \left(z^{\left(1\right)}_j,z^{\left(2\right)}_j\right)\right|
\leq C \left(M h\right)^{1/2},
\end{equation}
and for any non-negative integer $p$, $d=0$ or $1$ and $k\leq M$
\begin{equation}\label{e:zRR}
h \sumd\limits_{j=0}^{\infty}
\left|x_j\right|^p
\left|\dfrac{\partial^d}{\partial z^d}
\Rz \left(z^{\left(1\right)}_j,z^{\left(2\right)}_k\right)\right|^2
\leq C \left(1+\left|x_k\right|\right)^{p+d-3/2}.
\end{equation}
\end{prop}

P r o o f of Proposition~\ref{p:Sum_Int}. Since
$\left|z_j^{\left(1,2\right)} - x_j\right| = \Obig\left(h\right)$,
$\left|\Im \Rz \left(x,x\right)\right| \leq C
\left(1+\left|x\right|\right)^{-3/2}$ and derivatives of $\Rz$ are bounded
near the real line, we obtain that
\begin{equation*}
\left|\Im\Rz \left(z_j^{\left(1\right)},z_j^{\left(2\right)}\right)\right|
\leq 2C \left(1+\left|x_j\right|\right)^{-3/2}
\end{equation*}
for $n > n_0$ with some integer $n_0$. Hence,
\begin{equation*}
h \sumd\limits_{j=0}^{M}
\left|\Im \Rz \left(z^{\left(1\right)}_j,z^{\left(2\right)}_j\right)\right|
\leq C h \sumd\limits_{j=0}^{M} \left(1+\left|x_j\right|\right)^{-3/2}
\leq C.
\end{equation*}

The second statement can be checked in a similar way. The proof of the
third statement consists of several steps. First, we change $z_j$ by $x_j$
in \eqref{e:zRR}. The error of this change is a combination of sums of
higher derivatives with extra factors $h$. These sums are small, because
for $z_j$ far from $z_k$ these derivatives admit the exponential bound, and
for $z_j \sim z_k,$ in view of equation \eqref{e:Aeq} and restriction
$\left|z_k\right|\leq C n^{1/6} \log n,$ every two extra derivatives will
give us the sum as in \eqref{e:zRR} with the factor  of order $n^{-1/2}\log
n$. After the change of $z_j$ by $x_j,$ we obtain the sum which can be
estimated by the integral
\begin{equation*}
C \intd\limits_{0}^{\infty}x^p \left|\dfrac{\partial^d}{\partial z^d}
\Rz \left(x,z^{\left(2\right)}_k\right)\right|^2 dx,
\end{equation*}
because of the smoothness and exponential decreasing of $\Rz$. And finally,
the identity \eqref{e:prod} and Proposition~\ref{p:Rz_int} yield
\eqref{e:zRR}. We used the identity \eqref{e:prod} which is valid for real
$x$, but it remains valid for complex $x$ because the l.h.s and r.h.s of
the \eqref{e:prod} are entire functions equal at the real
line.\hfill\rule{0.5em}{0.5em}\medskip

{\bf Acknowledgement.}
 The author  is grateful to Prof. M.V. Shcherbina
 for  the problem statement and fruitful discussions.

\label{Poplavskyi.tex}

\end{document}